\documentclass[conference,compsoc]{IEEEtran}

\ifCLASSOPTIONcompsoc
  \usepackage[nocompress]{cite}
\else
  \usepackage{cite}
\fi

\ifCLASSINFOpdf
\else
\fi

\usepackage{xcolor}
\usepackage{graphicx} %
\usepackage{hyperref} %
\usepackage{subcaption}
\usepackage{multirow}
\usepackage{xspace}

\newcommand{\tool}[1]{RefChecker\xspace}
\newcommand{\mypara}[1]{\bigskip\noindent{\bf {#1}}}

\begin{document}
\title{Phantom References: Hallucinated Citations That Survive Peer Review\\ at Top‑Tier Conferences}

\author{\IEEEauthorblockN{Mark Russinovich\IEEEauthorrefmark{1},
Ram Shankar Siva Kumar\IEEEauthorrefmark{4},
Ahmed Salem\IEEEauthorrefmark{4}}
\IEEEauthorblockA{\IEEEauthorrefmark{1}Microsoft Azure}
\IEEEauthorblockA{\IEEEauthorrefmark{4}Microsoft}
}

\maketitle

\begin{abstract}
Large language models make it easy to produce scientific text that is polished, confident, and unsupported by real evidence.
When such text enters scholarly workflows, hallucination can become part of the archival record rather than merely a transient model error.
Measuring this risk through the prose of published papers is difficult: technical claims are contextual and often require expert judgment.
However, citations expose a narrower, more auditable surface.
A reference either resolves to a real scholarly work with compatible authorship, or it does not.

This paper measures \emph{citation hallucination} in peer-reviewed proceedings.
We define a conservative notion of hallucinated citation that counts identity-level failures: non-existent works and substantial author-list mismatches.
We explicitly exclude ordinary bibliographic drift, such as venue changes, year changes, publication-status updates, and minor name variants.
To audit citations at scale, we build \tool{}, a reference-verification pipeline that resolves bibliography entries against multiple bibliographic sources and escalates unresolved cases to web-search re-verification.
We apply \tool{} to accepted camera-ready papers from ICLR, ICML, NeurIPS, and USENIX Security.

Our results show that hallucinated citations have entered the archival record.
Reference-level rates are usually below one percent, but proceedings contain enough papers and references for these failures to become visible at the paper level.
In 2025, roughly one in twenty NeurIPS and USENIX Security papers contains at least two likely hallucinated academic-paper-like references under our strict definition.
We also observe post-ChatGPT increases in several venues highlighted by a high-count tail of papers with multiple (5+) failures in the same bibliography, and likely hallucinated citations even among award-winning papers.

These findings show that citation integrity is not reliably enforced by peer review alone, yet the problem is tractable: in one venue-scale scan, conference-scale auditing cost roughly four cents per paper.
We open-source \tool{} to support routine, reproducible citation verification before publication (\url{https://github.com/markrussinovich/refchecker}).
\end{abstract}

\IEEEpeerreviewmaketitle

\section{Introduction}
Large language models hallucinate~\cite{JLFYSXIBMF23, HYMZFWCPFQL25, ZLCCLFHZZCWLBSS25}. They fabricate quotations, invent sources, and produce confident, well-formed claims that have no base in the world. What began as a curiosity of early chatbots is now a recurring hazard in production: hallucinated content has surfaced in legal filings~\cite{mata2023avianca}, news reporting, code repositories~\cite{SWSMVJ25}, and technical documents, often in settings where downstream readers have neither the time nor the expertise to verify every claim. As LLMs are folded into the workflows that produce written knowledge, hallucination stops being only a model failure and starts becoming part of the public record.

Scientific writing is one of the workflows absorbing these tools the fastest. Researchers use LLMs to rephrase paragraphs, restructure arguments, summarize related work, and assemble bibliographies~\cite{LZWLJZCLHHYPMZ24}. A rough title, a partial memory of a paper, or a URL can become a polished BibTex entry in seconds~\cite{WW23, RC26}. On the surface, the resulting manuscript still looks like a paper: claims are supported, references are alphabetized, and the bibliography has the familiar shape of scholarly care. What has changed is the chain of evidence underneath, and with it the cost of placing a fabricated or misattributed reference into the scientific record.

At first glance this looks hard to measure. Checking an arbitrary technical claim in the body of a paper is essentially an open problem in automated fact-checking: claims are subtle, context-dependent, and tied to the surrounding argument, so even expert reviewers may disagree about the ground truth. Citations are different. A reference names a work and an authorship record: the cited work either exists in the bibliographic record or it does not, and the named authors either substantially match that record or they do not. That property makes the bibliography checkable in a way the surrounding text/claims are not.

We use such property to ask a pointed question of the published record: \emph{how often do accepted papers contain hallucinated citations?} By hallucinated citation we mean a reference that looks like an ordinary scholarly pointer but fails at the level of identity: either no corresponding work can be found, or the best-matching work has an author list that diverges substantially from the citation. We deliberately exclude ordinary bibliographic drift (changes in year, venue, or publication status, and minor name variants), which we record as citation-quality issues but never count as hallucination. The measure also makes no claim about whether the surrounding text was generated by an LLM or whether the cited result is true.

To answer the question at scale, we built a reference-verification pipeline, namely \tool{}, that checks citations against arXiv, Semantic Scholar, OpenAlex, CrossRef, DBLP, and the ACL Anthology, then escalates unresolved or suspicious cases to an LLM-driven deep web search before reaching a verdict. We apply this pipeline to accepted papers from top AI and security venues for which camera-ready PDFs are openly accessible through OpenReview or official conference proceedings pages. Unlike arbitrary preprints or generated text samples, these are manuscripts that passed peer review and entered the archival record.

\begin{figure*}[t]
    \centering
    \begin{subfigure}[t]{0.49\textwidth}
        \centering
        \includegraphics[width=\linewidth]{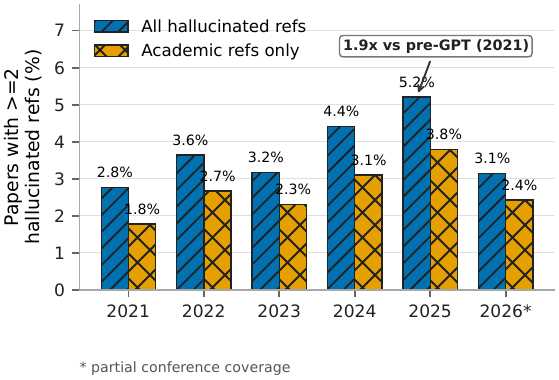}
        \caption{Affected-paper rate.}
        \label{fig:intro-teaser-affected-rate}
    \end{subfigure}\hfill
    \begin{subfigure}[t]{0.49\textwidth}
        \centering
        \includegraphics[width=\linewidth]{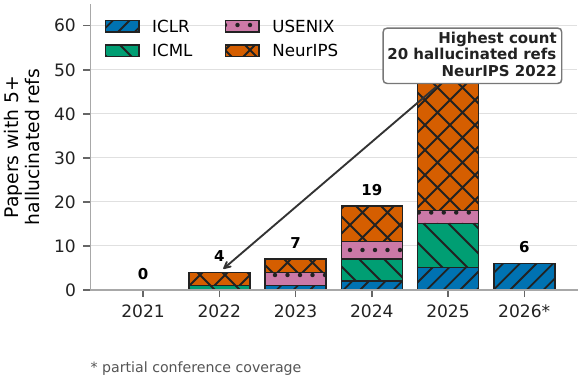}
        \caption{High-count tail.}
        \label{fig:intro-teaser-high-count-tail}
    \end{subfigure}
    \caption{Hallucinated citations in accepted conference papers. (a)~Even under our conservative definition, likely hallucinated references translate into visible paper-level prevalence once measured across full proceedings. (b)~The post-ChatGPT period shows a significantly higher-count tail: accepted papers with multiple likely hallucinated references in the same bibliography.}
    \label{fig:intro-teaser}
\end{figure*}

Our central finding turns on the choice of denominator. As \autoref{fig:intro-teaser} summarizes, the reference-level rate is usually below one percent. But proceedings contain thousands of papers and hundreds of thousands of references, and at that scale even a sub-percent failure rate leaves many accepted papers with hallucinated citations in their bibliographies. In 2025, restricting attention to academic-paper-like references, roughly one in twenty accepted papers at NeurIPS and USENIX Security contains at least two likely hallucinated references under our strict definition. Most affected papers contain only one such reference, which helps explain why the issue survives ordinary review; but the signal is not a scatter of isolated typos, because a visible tail of papers carries several (5+) hallucinated citations in a single bibliography. Nor is the problem confined to marginal papers: we observe likely hallucinated citations even among award-winning ones.

The timeline adds a second pattern. After the public release of ChatGPT, likely hallucinated-citation rates rise in several venue families and stay visible in recent proceedings, including the 2025 venue-years above. The series is not perfectly monotonic, and we do not treat the timing alone as proof that LLM-assisted writing caused the increase. But it is consistent with a change in authoring practice: bibliographic entries are now easier than ever to generate, complete, or repair from partial information, and the result can be syntactically polished while still failing to identify the work it purports to cite.

The problem has also become operationally urgent for conference governance.
ICLR~2026 states that papers with undisclosed or an erroneous LLM use that leads to false claims, misrepresentations, or hallucinated references can be desk rejected, while emphasizing that disciplinary action should rest on concrete evidence rather than detector scores alone~\cite{iclr2026llmPolicy}.
ICML~2026 similarly lists hallucinated references as grounds for desk rejection, and treats low-quality AI-generated content (``AI slop'') as interference with the peer-review process~\cite{icml2026peerReviewEthics}.
Public audits are already feeding into this governance shift: GPTZero reported $50$ ICLR~$2026$ submissions with at least one human-verified hallucinated citation among $300$ scanned submissions, after those submissions had each received $3$-$5$ expert reviews~\cite{gptzero2025iclr}.
These policies and reports reinforce the central point of our study: citation hallucination is no longer only an authoring artifact, but an enforceable publication-integrity problem that requires scalable, evidence-preserving verification.

Auditing this layer is no longer expensive. As one concrete example from our study, scanning a single venue's 3{,}703 accepted papers and 221{,}281 references cost approximately \$157, or roughly four cents per paper.\footnote{Cost reflects LLM API pricing as of May~2026; actual cost depends on provider rates, model selection, and the share of references that require web search.} We open-sourced \tool{} so that the same analysis can be reproduced and extended.

Given the sensitivity of these findings, we report them with deliberate restraint. We do not disclose the identity of any specific paper, author, or institution, and all results are presented in aggregate. This choice is not merely procedural. In discussions around preliminary versions of these results, researchers raised a concrete concern: public identification of affected papers could cause reputational harm disproportionate to what the audit can establish, especially when a citation failure may reflect a workflow breakdown rather than individual misconduct. \emph{The purpose of this work is therefore not to assign blame, but to expose a failure mode that current publication workflows are not designed to catch}: hallucinated citations that look valid on the page, pass through review, and yet point to no real work or to the wrong authorship record.

Hallucinated citations are at once subtle, plausible, and externally checkable, a combination that makes them both consequential and tractable. As LLMs become part of how science is written, preserving the integrity of the citation layer will require more than diligent authorship; it will require routine, automated verification before publication.

\noindent \textbf{Contributions.}
\begin{itemize}
    \item We define a conservative, externally auditable notion of hallucinated citation based on non-existent works and substantial author-identity mismatch, explicitly excluding ordinary venue, year, and publication-status drift.
    \item We build \tool{}, a scalable open-source reference-verification pipeline that combines multiple authoritative bibliographic sources with deep web-search re-verification for unresolved cases.

    \item We conduct a large-scale audit of likely hallucinated citations in accepted papers from top AI and security venues.
    \item We show that likely hallucinated citations survive peer review and appear in camera-ready proceedings, including award-winning papers and papers with multiple failures in the same bibliography.
    \item We quantify the post-ChatGPT increase, paper-level prevalence, error composition, and the high-count tail, and show that conference-scale auditing can be performed at low cost.
\end{itemize}

\section{Related Work}
\label{sec:related}

\mypara{Auditing hallucination.}
Closest to us are measurements of fabricated or misattributed references.
Zhao~et~al.~\cite{ZWSDGY26} audit $111$M references across $2.5$M
arXiv, bioRxiv, SSRN, and PubMed papers, but their corpus is dominated by
preprints rather than camera-ready proceedings.
Another concurrent study most comparable to ours is
GhostCite~\cite{GhostCite26}: $2.2$M citations from $56{,}381$ papers
across AI/ML and security venues ($2020$--$2025$), with $1.07\%$ of papers
carrying an invalid citation and a researcher survey
in which $76.7\%$ of reviewers report not checking references, consistent with our own findings.
GhostCite, however, is less conservative in selecting candidate hallucinations and would require more manual effort to filter them.
Other studies confirm the trend within single communities, e.g., Ansari~\cite{ansari26} on NeurIPS~2025
and Sakai~et~al.~\cite{SKW26} on the ACL conference family ($2024$ and $2025$).
Earlier reports likewise showed ChatGPT's tendency to fabricate citations~\cite{WW23}, as do recent GPTZero disclosures for ICLR~2026~\cite{gptzero2025iclr} and NeurIPS~2025~\cite{gptzero2026neurips}.

\mypara{Policy responses.}
Hallucinated references are now a publication-integrity concern: ICLR~$2026$
permits desk rejection for LLM use yielding hallucinated references
while insisting on concrete evidence over detector scores~\cite{iclr2026llmPolicy},
and ICML~$2026$ lists them among desk-reject grounds and treats ``AI slop'' as
interference with review~\cite{icml2026peerReviewEthics}.

\mypara{Fabrication under controlled prompting.}
A complementary line measures how often models fabricate references when
prompted directly.
Naser~\cite{naser26} audits $69{,}557$ model-generated citations across
ten deployed models and reports hallucination rates of $11.4\%$--$56.8\%$
depending on model and domain.
Agrawal~et~al.~\cite{ASMK24} further show that
models can often signal when a reference they produced is fabricated, and Rao
and Callison-Burch~\cite{RC26} show that pairing models with
deterministic retrieval sharply improves BibTex generation.
These studies characterize models in isolation; we instead measure what
survives peer review into the archival record.

\mypara{Citation-verification tools.}
Several systems target citation verification as deployable infrastructure.
CiteAudit~\cite{SSZSCY26} pairs a benchmark with a multi-agent
extract-retrieve-judge pipeline, and GhostCite's
CiteVerifier~\cite{GhostCite26} takes an approach similar to ours, resolving each reference against authoritative databases (arXiv, Semantic~Scholar, DBLP, and others) before any web search and escalating only suspicious cases to deep re-verification.

\mypara{Hallucination, attribution, and fact verification.}
More broadly, hallucination has been surveyed
extensively~\cite{JLFYSXIBMF23,HYMZFWCPFQL25,ZLCCLFHZZCWLBSS25} and
quantified by benchmarks such as TruthfulQA~\cite{truthfulqa},
HaluEval~\cite{halueval}, and FActScore~\cite{factscore},
building on faithfulness studies in
summarization~\cite{MNBM20,MLCHJS23}.
A parallel line of works makes generations checkable through attribution and grounded
citation, measuring whether a statement is supported by its source~\cite{RNLACDPTTR23}, evaluating verifiability in generative
search engines~\cite{LZL23}, and training models to emit inline
citations grounded in retrieved evidence~\cite{GYYC23,ZBLGLZCHDFL25}, on
the foundation of automated fact verification~\cite{TVCM18} and
alongside AI-text detection~\cite{KGWKMG23,MLKMF23}.

\mypara{Positioning.}
We are most comparable to Zhao~et~al.~\cite{ZWSDGY26} and
GhostCite~\cite{GhostCite26}.
We differ from these concurrent works by auditing camera-ready proceedings rather than
preprints, and from the latter on precision, paper-level structure
(affected-paper prevalence and a high-count tail), and reported per-paper cost.
Like both, we report only in aggregate and read the convergence of independent
measurements as evidence that the phenomenon is real and growing.

\section{Background and Definitions}
\label{sec:background}

\mypara{What a citation is.}
Scientific writing relies on a compact convention: a claim that draws on prior work is tied to a short, structured record that is at once an intellectual gesture and a pointer.
As a gesture, it credits the work on which an argument depends.
As a pointer, it gives a reader enough information (authors, title, venue, year, and identifiers) to recover the cited artifact from the public record.
The first role is social; the second is what makes a bibliography auditable, and it is the role that LLM-assisted authoring can quietly destabilize.

\mypara{Citation hallucination.}
We say a citation is \emph{hallucinated} when this pointer fails in a way that changes the identity of the cited work.
We mainly classify these errors in two classes:
The first is \emph{fabrication}: the cited work does not exist at all, no indexed artifact matches the title, and the listed authors, real or not, never produced anything matching it.
The second is \emph{author-identity corruption}: a real paper exists, but the citation assigns it to a substantially different set of authors.
We treat these as two faces of one phenomenon and count both, because from a reader's perspective the distinction is cosmetic: in either case the reference does not point to the work it claims to credit, the intellectual debt cannot be verified, and any argument built on the cited result inherits an unchecked premise.
What we do \emph{not} count is ordinary bibliographic noise (minor name variants, spelling mistakes, and venue, year, or publication-status changes as a manuscript moves from preprint to proceedings), unless the author evidence no longer matches at a substantial level.

\mypara{Why the citation layer.}
This narrow framing buys us something the surrounding prose rarely offers: a checkable claim.
A sentence in the body of a paper can be hard to judge even for an expert, since it may rest on modeling assumptions, unstated context, or a particular reading of prior work, and deciding whether it is ``hallucinated'' becomes a form of automated fact-checking.
A citation instead names an artifact and an authorship record, and bibliographic indices can usually tell us whether such a work exists and whether the cited authors match the public record.
Our test therefore asks neither whether the cited claim is true, nor whether the venue or year is the most current, but only a narrower question: does the bibliography point to a real work by the named authors?
That is what lets us measure citation hallucination at conference scale without turning the study into a subjective review of each paper's technical argument.

\section{Methodology}
\label{sec:method}

\mypara{Measurement approach.}
We measure hallucinated citations at the reference level: each bibliography entry is treated as a claim that should resolve to a compatible source record.
This scope is deliberately narrower than paper authorship or claim verification.
We do not decide whether a paper was written by an LLM, whether a cited claim is semantically supported, or whether every bibliographic discrepancy is important.
For each citation, \tool{} records the source candidates it finds, the metadata fields it compares, and the final status assigned by the pipeline.
This makes each counted hallucination traceable to the supporting source lookup and metadata comparison.

\begin{figure*}[t]
    \centering
    \includegraphics[width=\textwidth]{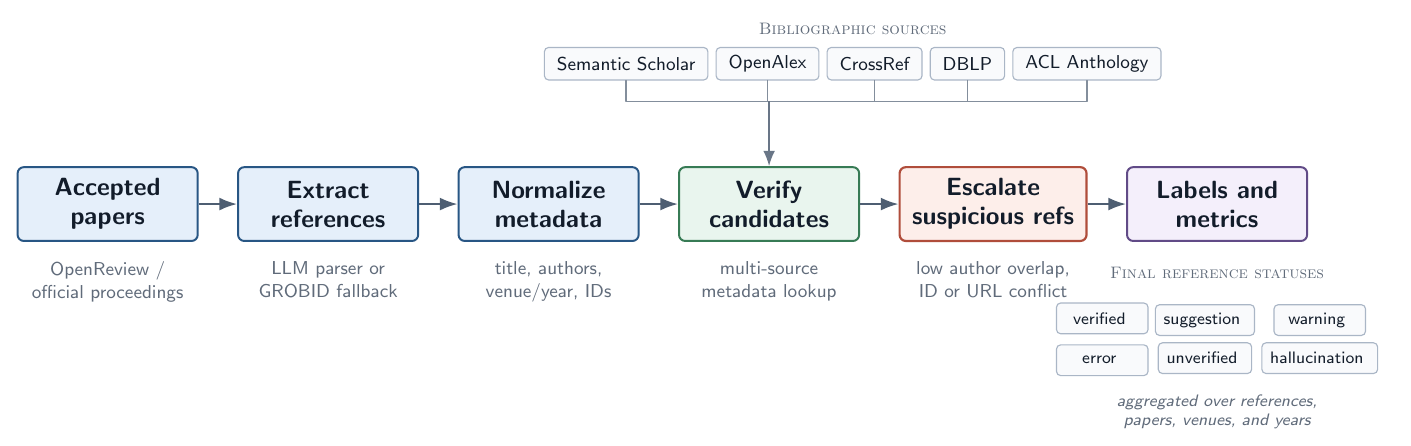}
    \caption{Measurement pipeline. Papers are parsed into references, normalized, checked against bibliographic sources (Semantic Scholar, OpenAlex, CrossRef, DBLP, ACL Anthology), escalated when identity evidence is suspicious, and aggregated into one of six final reference statuses. Lower-level issue values, infrastructure error types, and hallucination verdicts are defined in \autoref{tab:refchecker-vocabulary}.}
    \label{fig:methodology-pipeline}
\end{figure*}

\mypara{Corpus construction.}
Our corpus consists of accepted papers from ICLR, ICML, NeurIPS, and USENIX Security whose camera-ready PDFs are openly accessible through OpenReview or official conference proceedings pages.
For OpenReview venues, we collect accepted-paper records and their linked PDFs.
For USENIX Security, we collect PDFs from the official proceedings pages.
We exclude papers for which no camera-ready PDF can be retrieved from these sources, and we do not use author-hosted copies as substitutes.
This restriction ensures that every analyzed manuscript is tied to a venue-controlled or platform-controlled publication record.

\mypara{Extraction and normalization.}
\tool{} supports PDFs, Latex files, BibTex files, plain text, OpenReview URLs, arXiv identifiers, URLs, and pre-extracted bibliographies.
In our measurement, the primary input is the camera-ready PDF.
The tool extracts the bibliography and parses each reference into fields such as title, authors, year, venue, DOI, arXiv identifier, and URL.
When configured with an extraction model, \tool{} uses LLM-assisted parsing for noisy bibliographies (for this work we use Gemini 3.1 Flash Lite); otherwise, PDFs can fall back to GROBID when available.
The original reference string is preserved alongside the parsed fields so downstream verification can be audited when parsing is noisy.
Before matching, titles, authors, identifiers, and venue/year fields are normalized to absorb harmless formatting variation without broadening the hallucination definition.

\mypara{Verification sources.}
Each normalized reference is checked against multiple bibliographic sources, including Semantic Scholar, OpenAlex, CrossRef, DBLP, and the ACL Anthology.
DOI evidence, arXiv identifiers, and cited URLs are also used when present.
The verifier searches for candidate works and compares the cited title, authors, year, venue, and identifiers against the best available records.
Using multiple sources is necessary because coverage and metadata quality vary significantly across venues, years, and publication types.
When sources disagree, \tool{} preserves the evidence rather than treating any single source as absolute ground truth.

\mypara{Hallucination escalation.}
Most references do not require deep hallucination checking.
\tool{} first applies deterministic filters: a reference is escalated when it cannot be verified, when author overlap falls below the tool's default $60\%$ threshold for references with at least three authors, when a DOI or arXiv identifier resolves to a different work, or when a cited URL fails or points elsewhere.
Minor year or venue variation alone is not sufficient for escalation.
Escalated references are sent to an LLM with web-search capability for verification.
The search task is constrained: the model must look for a dedicated source page for the cited work, not merely another paper that repeats the citation.
If the model finds a candidate source, \tool{} re-verifies the cited metadata against that source before clearing the reference or leaving it suspicious.
Only references that remain unsupported, or whose best available source conflicts substantially with the cited authorship record, are counted as likely hallucinated.

\begin{table*}[t]
    \centering
    \scriptsize
    {\renewcommand{\arraystretch}{1.08}
    \begin{tabular}{@{}p{0.16\textwidth}p{0.23\textwidth}p{0.53\textwidth}@{}}
        \hline
        \textbf{Kind} & \textbf{Value} & \textbf{Meaning in our measurement} \\
        \hline
        \multirow{6}{=}{\centering\textbf{Final status}}
            & \texttt{verified} & The reference resolves to a compatible scholarly source. \\
            & \texttt{error} & A substantive mismatch remains, such as title, author, DOI, URL, or identifier conflict. \\
            & \texttt{warning} & A weaker discrepancy remains, such as year or venue variation that may reflect publication-version drift. \\
            & \texttt{suggestion} & An informational improvement is available, such as adding a missing DOI, arXiv ID, or URL. \\
            & \texttt{unverified} & The configured sources could not confirm the reference. \\
            & \texttt{hallucination} & The hallucination assessment concludes that the reference is likely fabricated or substantially conflicts with the cited authorship record. \\
        \hline
        \multirow{10}{=}{\centering\textbf{Issue value}}
            & \texttt{title} & The cited title conflicts with the matched source or identifier target. \\
            & \texttt{author} & The cited authors conflict with the matched source; severe author mismatch is part of our hallucination definition. \\
            & \texttt{year} & The cited year differs from the matched source. \\
            & \texttt{venue} & The cited venue differs from the matched source. \\
            & \texttt{doi} & The cited DOI is missing, inconsistent, or resolves to a different work. \\
            & \texttt{arxiv\_id}, \texttt{arxiv} & The cited arXiv identifier is inconsistent with the matched work or points elsewhere. \\
            & \texttt{url} & The cited URL is broken, missing, or points to a different object. \\
            & \texttt{unverified} & No configured source confirms the reference. \\
            & \texttt{multiple} & The consolidated record contains more than one issue value. \\
            & \texttt{info}, \texttt{suggestion\_*} & Informational or recommendation-only entries, such as \texttt{suggestion\_doi} or \texttt{suggestion\_arxiv\_url}. \\
        \hline
        \centering\textbf{Operational value}
            & \texttt{api\_failure}, \texttt{processing\_failed}, \texttt{check\_failed}, \texttt{unexpected\_error}, \texttt{unknown} & Infrastructure or execution failures retained for auditing; these are not hallucination labels. \\
        \hline
        \centering\textbf{Hallucination verdict}
            & \texttt{LIKELY}, \texttt{UNLIKELY}, \texttt{UNCERTAIN} & The deep-check verdict; only \texttt{LIKELY} verdicts that survive metadata re-verification are counted as likely hallucinated. \\
        \hline
    \end{tabular}}
    \caption{\tool{} output vocabulary used in the audit. }
    \label{tab:refchecker-vocabulary}
\end{table*}

\mypara{Labels used in the audit.}
\autoref{tab:refchecker-vocabulary} distinguishes three layers that are easy to conflate.
First, final statuses describe how a reference is reported after verification.
Second, issue values describe the evidence that led to that status.
Third, hallucination verdicts describe the outcome of deep checking for suspicious references.
Our aggregate hallucination measurements use only likely hallucinated references: references with a positive hallucination outcome after re-verification.
Errors, warnings, suggestions, and unverified references can inform triage, but they are not automatically counted as hallucinations.
Uncertain cases are retained in the audit log and excluded from positive hallucination counts unless explicitly stated otherwise.

\mypara{Use of LLMs.}
LLMs are used as components of the measurement pipeline, not as the sole authority for citation validity.
They help parse noisy bibliographies and perform targeted web searches for suspicious references.
The final decision is constrained by external evidence: database matches, identifiers, discovered source pages, and re-verification of title and authorship metadata.
This design reduces but does not eliminate model risk.
LLMs can miss obscure sources, over-trust low-quality pages, or return incomplete metadata; for that reason, the pipeline treats LLM search as an escalation and evidence-gathering step rather than as a direct labeler.

\mypara{Limitations.}
PDF extraction noise, incomplete or stale database records, and the behavior of the configured LLM provider can each produce false positives and false negatives, so we read the measured rates as an auditable estimate of citation-layer hallucination rather than an exact count.
We later present a more detailed limitation discussion in~\autoref{sec:limitations}.

\section{Evaluation}
\label{sec:results}

We ran the pipeline over every accepted paper we could retrieve from four venue families: ICLR ($2021$--$2026$), together with ICML, NeurIPS, and USENIX Security ($2021$--$2025$).
The corpus is $48{,}095$ accepted papers and $2{,}614{,}992$ extracted references, and the years it spans straddle the public release of ChatGPT.
Every rate we report, unless stated otherwise, is computed on that corpus.
One pattern drives this section: the same failures look very different depending on the denominator used to count them.
As a fraction of all references, hallucinated citations are rare (under one percent in almost every venue-year) and easy to dismiss.
As a fraction of accepted papers they are not rare at all, because a full proceedings holds enough references that even a sub-one-percent rate reaches a large number of distinct papers.
The rest of the section follows from that gap: what the failures actually are, where they concentrate, and why we believe peer review does not already catch them.

For the remainder of this section, a \emph{likely hallucinated reference} is the strict residue of the pipeline's verdict: either no matching work can be found, or the best match has an author list that diverges substantially from the citation.
Year, venue, and publication-status drift are recorded but \emph{not} counted, so a reference whose authors and title round-trip against the bibliographic record, even after a preprint becomes a proceedings paper, is treated as correct.
The numbers below therefore understate any broader notion of citation noise and are best read as a lower bound.

\subsection{Metrics}
We aggregate at four levels: the individual reference, the paper, the venue, and the year.
At the reference level we report the likely hallucinated-reference rate under two denominators.
The all-reference rate keeps the entire bibliography in the denominator, including web pages, standards, RFCs, software, datasets, and other non-paper entries.
The academic-paper-like rate restricts the denominator to references that identify as scholarly works.

At the paper level we report affected-paper prevalence at two thresholds: the share of accepted papers with at least one likely hallucinated reference, and the share with at least two.
The two thresholds trade coverage against confidence.
At least one captures raw exposure, but it is the most sensitive to false positives, because a single mistaken flag is enough to count a paper.
At least two is a more conservative reading: two different references in the same paper have to fail before it is counted, so an isolated verification error no longer tips a paper over the line.

Among affected papers we report the full distribution of per-paper counts, so that a heavy tail cannot hide behind a low average.
We treat that tail as a metric in its own right: the prevalence of papers with five or more likely hallucinated references, the fraction of all likely hallucinated references those papers contribute, and the single-paper maximum in each venue family.
The tail is also where our confidence is highest: the more failures a single bibliography accumulates, the less plausible it is that every one of them is a false positive, so the five-or-more papers are the cases we can most safely call genuine.

Finally, we report error composition.
Each likely hallucinated reference is assigned to one identity-level failure mode: an unresolved reference for which no matching work was found, a substantial author-list mismatch against the best-matching work, or a multi-field failure combining at least two such inconsistencies.
Ordinary venue, year, and publication-status drift is logged separately and never enters any rate above.
Where a venue publishes peer-review metadata through OpenReview, we additionally join each paper's likely hallucinated-reference count to its mean reviewer rating, its acceptance tier, and its declared primary research area.

\begin{figure*}[t]
    \centering
    \begin{subfigure}[t]{0.49\textwidth}
        \centering
        \includegraphics[width=\linewidth]{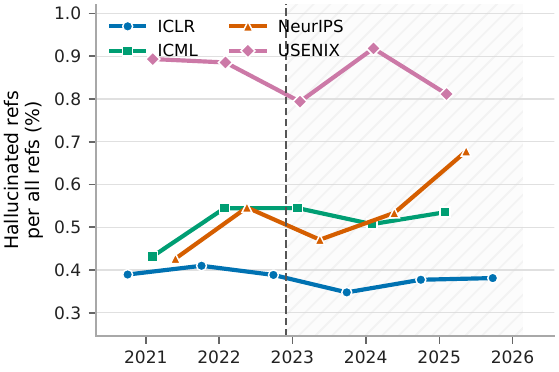}
        \caption{All extracted references.}
        \label{fig:reference-rate-all}
    \end{subfigure}\hfill
    \begin{subfigure}[t]{0.49\textwidth}
        \centering
        \includegraphics[width=\linewidth]{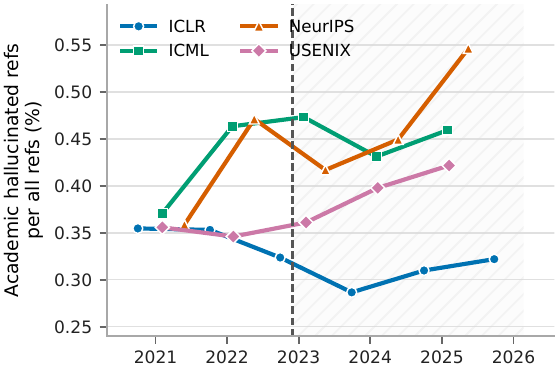}
        \caption{Academic-paper-like references only.}
        \label{fig:reference-rate-academic}
    \end{subfigure}
    \caption{Reference-level rate of likely hallucinated references by venue-year.
    (a) all extracted references; (b) restricted to academic-paper-like references.
    The shaded region marks proceedings whose camera-ready window opens after the public release of ChatGPT.
    The $2026$ point is partial: ICLR only.}
    \label{fig:reference-rate-timeline}
\end{figure*}

\subsection{Findings}

\mypara{The reference-level rate is small but misleading.}
\autoref{fig:reference-rate-timeline} answers the most literal version of the question.
Across our corpus, the share of likely hallucinated references stays below one percent in essentially every venue-year.
For the latest complete year, $2025$, the all-reference rate is $0.38\%$ at ICLR, $0.54\%$ at ICML, $0.68\%$ at NeurIPS, and $0.81\%$ at USENIX Security; restricting the denominator to academic-paper-like references shifts these to $0.31\%$, $0.46\%$, $0.55\%$, and $0.42\%$, respectively.

These numbers are easy to misread.
A sub-one-percent rate sounds negligible for a single bibliography.
A conference proceedings is not a single bibliography.
ICLR~$2025$ alone contains $221{,}281$ extracted references across $3{,}703$ accepted papers (that were successfully extracted by our tool); even $0.38\%$ of that is $835$ flagged references in $692$ distinct papers.
The post-ChatGPT trend does not show clearly at this level, where false-positive noise dominates; it sharpens once we focus on papers with several likely hallucinated references, which are far less likely to be false positives.

\begin{figure*}[t]
    \centering
    \begin{subfigure}[t]{0.49\textwidth}
        \centering
        \includegraphics[width=\linewidth]{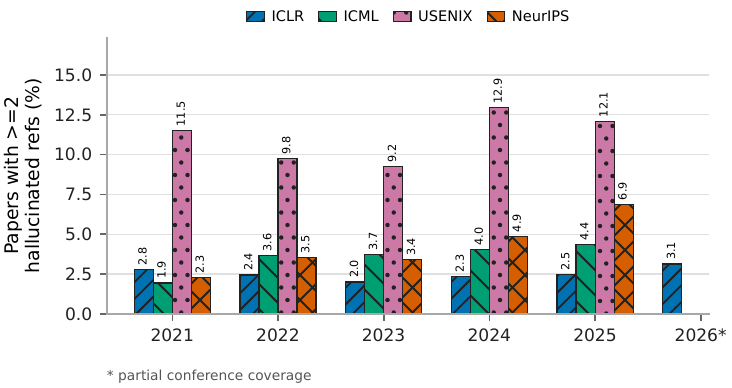}
        \caption{All extracted references.}
        \label{fig:paper-prevalence-ge2a}
    \end{subfigure}\hfill
    \begin{subfigure}[t]{0.49\textwidth}
        \centering
        \includegraphics[width=\linewidth]{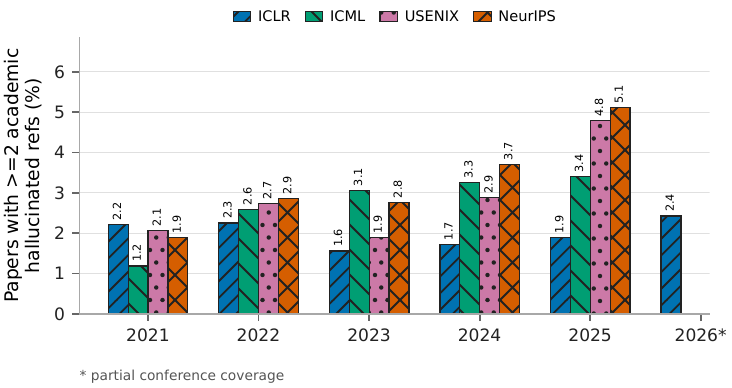}
        \caption{Academic-paper-like references only.}
        \label{fig:paper-prevalence-ge2b}
    \end{subfigure}
    \caption{Share of accepted papers with at least two likely hallucinated references, by venue and year.
    (a) over all extracted references; (b) over academic-paper-like references only.}
    \label{fig:paper-prevalence-ge2}
\end{figure*}

\mypara{Per paper, the failure is common.}
The paper-level view changes the picture.
At least one likely hallucinated reference appears in $18.7\%$ of accepted ICLR~$2025$ papers, $23.3\%$ at ICML~$2025$, $26.2\%$ at NeurIPS~$2025$, and $34.9\%$ at USENIX Security~$2025$.
Roughly one in four accepted papers at NeurIPS~$2025$, and one in three at USENIX Security~$2025$, contains at least one reference that the pipeline cannot reconcile against any work in its bibliographic backends.

Raising the threshold to two does not flatten the signal.
\autoref{fig:paper-prevalence-ge2a} reports the share of accepted papers with at least two likely hallucinated references over all extracted references.
For $2025$ this is $2.5\%$ at ICLR, $4.4\%$ at ICML, $6.9\%$ at NeurIPS, and $12.1\%$ at USENIX Security.
Restricting to academic-paper-like references (\autoref{fig:paper-prevalence-ge2b} lowers the numbers but not to zero: $1.9\%$ at ICLR, $3.4\%$ at ICML, $5.1\%$ at NeurIPS, and $4.8\%$ at USENIX Security.
Roughly one in twenty accepted papers at NeurIPS and USENIX Security in $2025$ carries at least two references likely hallucinated, i.e., whose scholarly identity does not survive external verification, even after we strip out non-paper citations.

These rates rise in the post-ChatGPT window in three of the four venue families, and the ICLR~$2026$ point ($3.1\%$ on all references; $2.4\%$ on academic-paper-like) is the highest ICLR rate we observe.

\begin{figure}[t]
    \centering
    \includegraphics[width=\columnwidth]{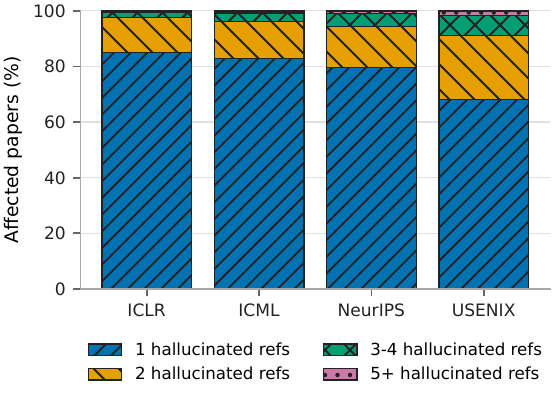}
    \caption{Distribution of likely hallucinated-reference counts among \emph{affected} papers.
    Bars are normalised within each conference family.}
    \label{fig:affected-paper-concentration}
\end{figure}

\mypara{Affected bibliographies are mostly clean, with a long tail.}
\autoref{fig:affected-paper-concentration} shows what an affected paper typically contains.
Most carry a single bad pointer in an otherwise clean bibliography: among affected papers, exactly one likely hallucinated reference accounts for $87\%$ at ICLR, $81\%$ at ICML, $74\%$ at NeurIPS, and $65\%$ at USENIX Security in $2025$.
Two such references account for another $9$--$27\%$ depending on venue.
These distributions argue against framing affected papers as compromised manuscripts.
They resemble ordinary submissions that happen to carry one or two flawed citations; we suspect many are not LLM hallucinations at all, but accidental typos or false positives from our own pipeline.

The tail matters more for camera-ready governance.
USENIX Security is the heaviest at the top end: $34.6\%$ of affected USENIX Security papers in $2025$ contain at least two likely hallucinated references, and $2.0\%$ contain at least five.
NeurIPS~$2025$ shows the same pattern at higher absolute counts: $1{,}545$ affected papers, $407$ of them with at least two flagged references, and $30$ with at least five.
A single bad pointer is the typical failure mode, but the more flagged references a paper carries, the more likely those references are genuine hallucinations.

\begin{figure*}[t]
    \centering
    \begin{subfigure}[t]{0.49\textwidth}
        \centering
        \includegraphics[width=\linewidth]{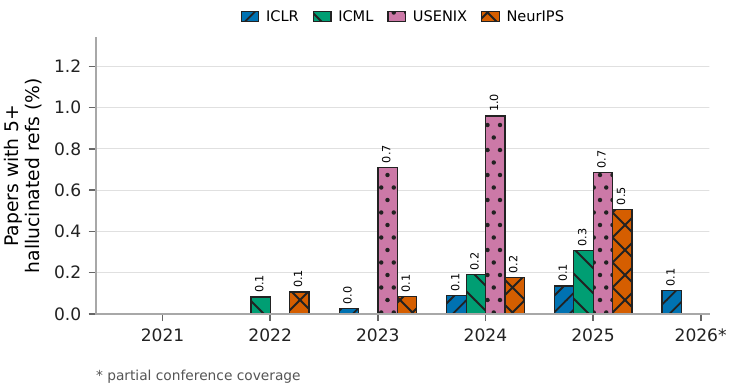}
        \caption{Papers with $\geq 5$ likely hallucinated references (\%).}
        \label{fig:ge5-prevalence}
    \end{subfigure}\hfill
    \begin{subfigure}[t]{0.49\textwidth}
        \centering
        \includegraphics[width=\linewidth]{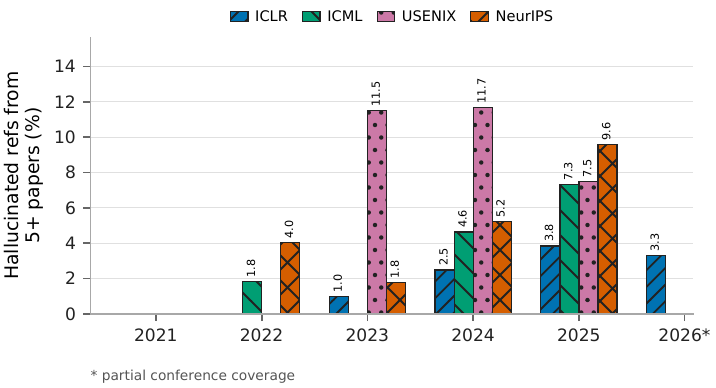}
        \caption{Share of all likely hallucinated references coming from those high-count papers (\%).}
        \label{fig:ge5-likely-share}
    \end{subfigure}
    \caption{Concentration of likely hallucinated references in the high-count tail.
    Papers with five or more flagged references are rare, but they account for a disproportionate share of all hallucinated references in the venue-years where they appear.}
    \label{fig:high-count-tail}
\end{figure*}

\begin{figure}[t]
    \centering
    \includegraphics[width=\columnwidth]{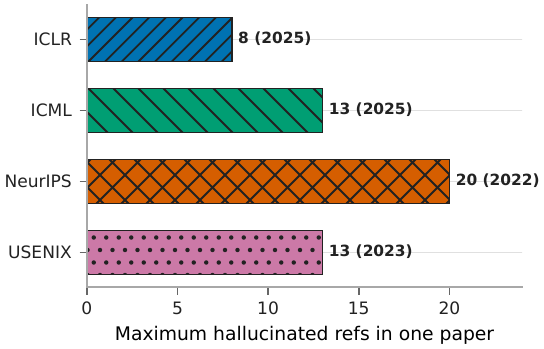}
    \caption{Single-paper maximum number of likely hallucinated references observed in each conference family across our corpus.}
    \label{fig:max-by-conference}
\end{figure}

\mypara{The tail is rare, but it concentrates risk.}
\autoref{fig:high-count-tail} isolates the high-count tail: accepted papers with at least five likely hallucinated references in a single bibliography.
On a paper-rate basis these are rare in every venue-year ($\leq 1\%$, often zero pre-$2022$).
On a reference-count basis they are not rare at all.
At USENIX Security~$2024$ the high-count tail accounts for $11.7\%$ of all flagged references; at NeurIPS~$2025$, $9.6\%$; at USENIX Security~$2025$, $7.5\%$; at ICML~$2025$, $7.3\%$.
A camera-ready check that flagged only papers with five or more failures would still be intercepting a non-trivial fraction of the underlying hallucinated citations.

The per-paper maxima sharpen the same point.
\autoref{fig:max-by-conference} reports the highest single-paper count we observed in each family across our entire corpus: $20$ at NeurIPS, $13$ at both ICML and USENIX Security, and $8$ at ICLR.
A bibliography with $20$ likely hallucinated references is not a borderline case.
It is a manuscript whose related-work claims, taken as a whole, might no longer align with the published scholarly record.

\begin{figure}[t]
    \centering
    \includegraphics[width=\columnwidth]{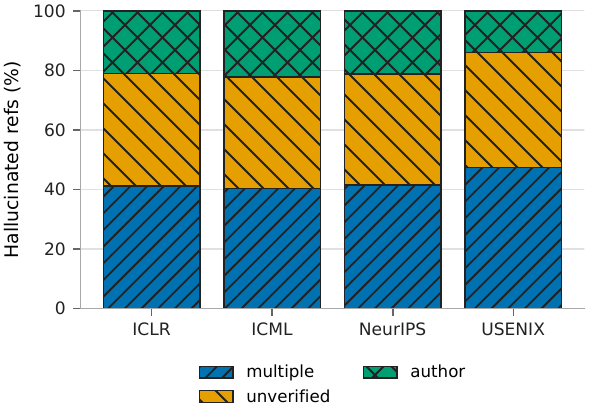}
    \caption{Composition of likely hallucinated references by failure mode.
    \emph{Unverified}: no corresponding work could be found.
    \emph{Author}: substantial author-list mismatch against the best-matching work.
    \emph{Multiple}: at least two simultaneous identity-level inconsistencies; ordinary venue or year drift alone is excluded.}
    \label{fig:error-type-composition}
\end{figure}

\mypara{The failures are identity errors, not bibliographic drift.}
A natural objection at this point is that the rates measure citation noise rather than hallucination: an arXiv year written instead of a venue year, a transliterated author name, an outdated URL.
\autoref{fig:error-type-composition} answers this directly.
Across every venue family, the likely hallucinated references decompose into three identity-level buckets: unresolved references, multi-field failures, and substantial author-list mismatches.

\begin{figure}[t]
    \centering
    \includegraphics[width=\columnwidth]{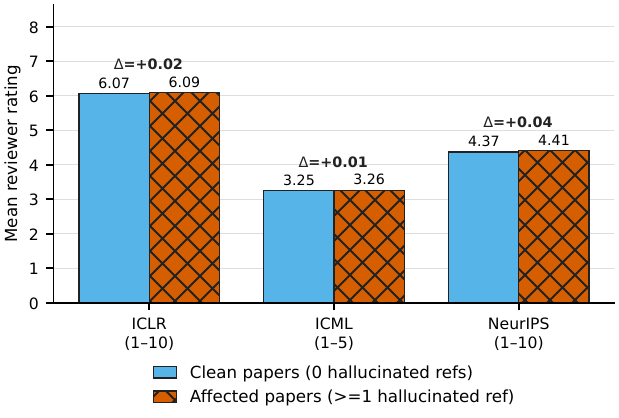}
    \caption{Mean reviewer rating, clean vs.\ affected papers, computed on OpenReview-covered accepted papers only (some NeurIPS and ICML papers/years are excluded due to non-public reviews).
    Venue scales differ ($1$--$10$ for ICLR/NeurIPS, $1$--$5$ for ICML).}
    \label{fig:review-blind-spot}
\end{figure}

\begin{figure}[t]
    \centering
    \includegraphics[width=\columnwidth]{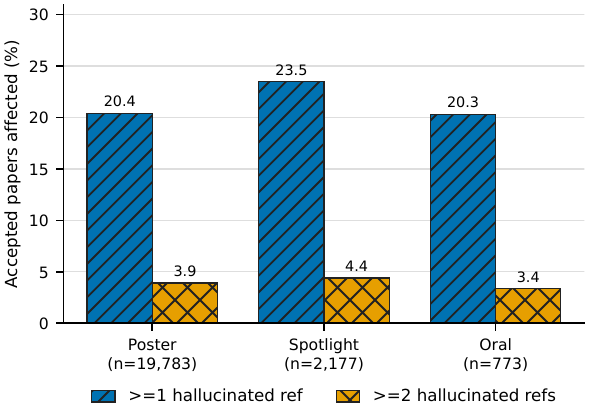}
    \caption{Hallucinated-reference prevalence by acceptance tier (OpenReview-covered accepted papers only).}
    \label{fig:acceptance-tier}
\end{figure}

\mypara{Peer review does not catch hallucinated citations.}
For the OpenReview-covered subset of our corpus, ICLR~$2021$/$2023$-$2026$, ICML~$2025$, and NeurIPS~$2025$, totalling $22{,}764$ accepted papers with public reviews, we can join each paper's hallucination count to its mean reviewer rating and to its decision tier.
\autoref{fig:review-blind-spot} compares the mean reviewer rating received by clean papers (zero likely hallucinated references) and affected papers (at least one).
The difference is almost negligible: $\Delta = +0.02$ at ICLR (scale 1--10), $+0.005$ at ICML (scale 1--5), and $+0.04$ at NeurIPS (scale 1--10) on each venue's own rating scale.
If anything, affected papers are rated marginally \emph{higher} than clean ones.
Whatever signals reviewers extract from a paper, ``does this bibliography hold up to external verification'' is not one of them.

The same blind spot shows up if we slice by acceptance tier instead of by rating (\autoref{fig:acceptance-tier}).
Posters are affected at $20.4\%$, spotlights at $23.5\%$, and orals at $20.3\%$.
Acceptance into the most selective tier does not lower hallucinated-citation prevalence.
This is consistent with the rating result: reviewers do not appear to be downgrading papers for bibliographic failures, so promotion to spotlight or oral status is uncorrelated with the bibliography's verification fate.
The recent governance shifts at ICLR~$2026$ and ICML~$2026$, which explicitly list hallucinated references as grounds for desk rejection, are responding to exactly this gap.

\emph{To be clear: we do not blame the reviewers, whose work is voluntary and who already contend with a sharp rise in submissions in recent years. We believe automated tools like ours should be provided and used by conferences to help reviewers by flagging potential citation problems that can then be verified manually.}

\begin{figure}[t]
    \centering
    \includegraphics[width=\columnwidth]{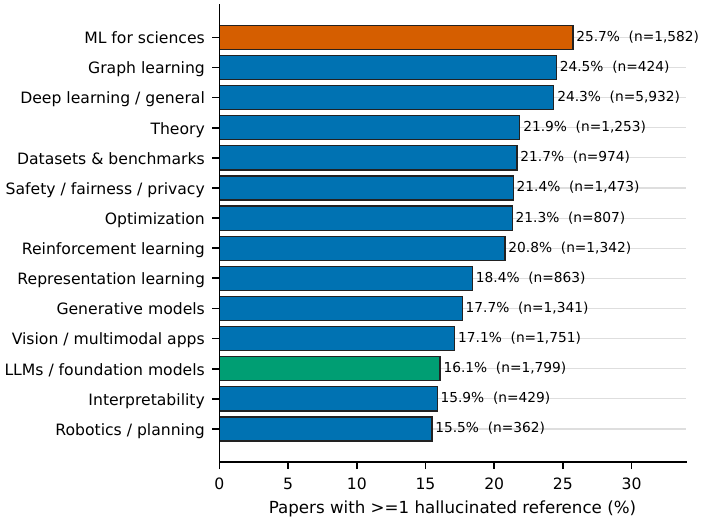}
    \caption{Prevalence of likely hallucinated references by primary research area, computed on OpenReview-covered accepted papers with declared primary areas ($n \geq 100$ per area).}
    \label{fig:area-prevalence}
\end{figure}

\mypara{The exposure is broad across research areas.}
The OpenReview metadata also tags each paper with a primary research area.
\autoref{fig:area-prevalence} reports affected-paper prevalence across the fourteen areas with at least $100$ papers in our covered subset.
Every area sits in a narrow band between $15.5\%$ and $25.7\%$.
The highest is \emph{ML for sciences} at $25.7\%$, followed by \emph{graph learning} ($24.5\%$) and \emph{deep learning / general} ($24.3\%$); the lowest are \emph{robotics / planning} ($15.5\%$), \emph{interpretability} ($15.9\%$), and \emph{LLMs / foundation models} ($16.1\%$).

These results suggest two things.
First, the problem cannot be reduced to a particular community's authoring practices.
Second, the community most fluent in LLMs (authors working on LLMs and foundation models) appears slightly less prone to the failure mode, possibly by checking references more carefully, possibly by using narrower tooling.
Overall, however, no area is exempt.

\begin{figure}[t]
    \centering
    \includegraphics[width=\columnwidth]{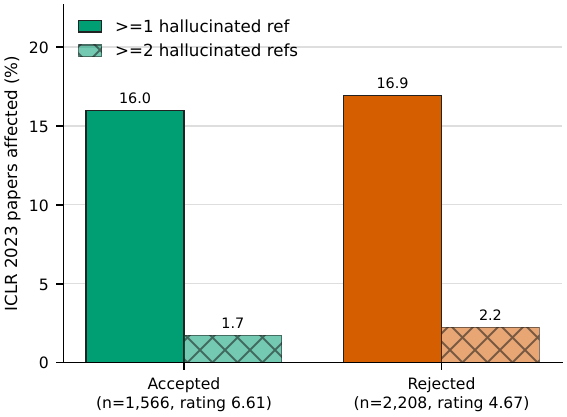}
    \caption{Hallucinated-reference prevalence in accepted vs.\ rejected submissions (ICLR~$2023$), with mean reviewer ratings shown in the $x$-axis labels.}
    \label{fig:iclr23-natural}
\end{figure}

\mypara{Accepted vs.\ rejected at ICLR~$2023$.}
We collected all publicly available rejected papers from OpenReview for ICLR~$2023$: $2{,}208$ \emph{rejected} submissions, in addition to the $1{,}566$ accepted ones. Running the pipeline over both groups (\autoref{fig:iclr23-natural}) yields a natural comparison for citation hallucination.
Accepted papers receive a mean reviewer rating of $6.61$ and contain at least one likely hallucinated reference $16.0\%$ of the time; while rejected papers receive a mean rating of $4.67$, almost two full points lower, yet are also affected $16.9\%$ of the time.
The gap in bibliographic integrity between the two groups is negligible, even though the gap in reviewer assessment is large.

Two conclusions follow.
First, the hallucinated-citation rate is driven by authoring practice rather than by overall paper quality: a paper that reviewers judge to be much weaker is not, in expectation, much more likely to carry hallucinated citations.
Second, reviewers' rating signal is essentially orthogonal to bibliographic integrity.
Both observations point in the same direction as the earlier results: the signal exists in the published record, but it is invisible to the process that decides what gets published.

\mypara{Summary.}
Taken together, the results describe a specific failure mode rather than a general indictment.
At the reference level the rate is small, and we have been deliberate about not inflating it.
At the paper level the prevalence is large: roughly one in four accepted papers at NeurIPS~$2025$ contains at least one likely hallucinated reference, and roughly one in twenty contains at least two even after stripping non-paper citations.
Affected papers usually contain only one such reference, but a heavy tail contributes a disproportionate share of all failures and produces individual cases with double-digit counts.
Reviewer ratings and acceptance tiers carry essentially no information about whether a paper's bibliography survives verification, and the rate barely changes between accepted and rejected submissions whose ratings differ by almost two points.

This is a problem that automated pre-publication verification is well suited to solve and that human peer review, in its current form, demonstrably is not.
The new ICLR~$2026$ and ICML~$2026$ rules treat hallucinated references as a desk-reject signal; the numbers above show that the signal is there in the proceedings, is large enough to matter, and is not being captured by anything else in the pipeline.

\makeatletter
\@ifundefined{definecolor}{%
    \newcommand{\exPlainCard}[5]{%
        \par\smallskip\noindent
        {\setlength{\fboxsep}{5pt}\fbox{%
        \begin{minipage}{\dimexpr\linewidth-2\fboxsep-2\fboxrule\relax}
        \scriptsize
        \textbf{#1}\\[0.25em]
        \begin{tabular}{@{}p{0.13\linewidth}p{0.81\linewidth}@{}}
        \textbf{Cited} & #2 \\
        \textbf{Flagged} & #3 \\
        \textbf{Verified} & #4 \\
        \textbf{Decision} & #5 \\
        \end{tabular}
        \end{minipage}}}
        \par\smallskip\noindent
    }
    \newcommand{\exTPCard}[5]{\exPlainCard{TP: #1}{#2}{#3}{#4}{#5}}
    \newcommand{\exFPCard}[5]{\exPlainCard{FP: #1}{#2}{#3}{#4}{#5}}
}{%
    \definecolor{exTPFill}{RGB}{238,247,241}
    \definecolor{exTPBorder}{RGB}{55,123,85}
    \definecolor{exFPFill}{RGB}{253,240,237}
    \definecolor{exFPBorder}{RGB}{176,77,59}
    \newcommand{\exColorCard}[7]{%
        \par\smallskip\noindent
        {\setlength{\fboxsep}{5pt}\fcolorbox{#2}{#1}{%
        \begin{minipage}{\dimexpr\linewidth-2\fboxsep-2\fboxrule\relax}
        \scriptsize
        \textbf{#3}\\[0.25em]
        \begin{tabular}{@{}p{0.13\linewidth}p{0.81\linewidth}@{}}
        \textbf{Cited} & #4 \\
        \textbf{Flagged} & #5 \\
        \textbf{Verified} & #6 \\
        \textbf{Decision} & #7 \\
        \end{tabular}
        \end{minipage}}}
        \par\smallskip\noindent
    }
    \newcommand{\exTPCard}[5]{\exColorCard{exTPFill}{exTPBorder}{TP: #1}{#2}{#3}{#4}{#5}}
    \newcommand{\exFPCard}[5]{\exColorCard{exFPFill}{exFPBorder}{FP: #1}{#2}{#3}{#4}{#5}}
}
\makeatother

\section{Discussion}
\label{sec:discussion}

Our measurement establishes that hallucinated citations have entered the archival record and that peer review does not reliably remove them.
We now turn to what that finding implies for how the problem should be defined, adjudicated, and governed.

\mypara{Separating hallucination from ordinary error is genuinely hard.}
The single most consistent lesson from our audit is that an automated flag is a starting point, not a verdict.
Across the references we manually verified, most flags were not hallucinations: mangled author strings, truncated titles, and references corrupted during PDF extraction.
The flags that survived scrutiny shared a sharper property, the reference failed at the level of \emph{identity}: no corresponding work could be found, or a real title was bound to an authorship record unrelated to the real one.
We use a small set of adjudicated examples below to make this boundary concrete; each is reported without deanonymizing the citing paper.

\mypara{What a true hallucination looks like.}
The clearest true positives stayed superficially plausible while breaking the identity of the cited work.
The most fundamental case is a reference that resolves to nothing at all: a well-formed title, plausible authors, and a real-sounding venue that together correspond to no work in any source we checked.

\exTPCard{No such work}
    {``Applying Novelty Detection Techniques for Quality Assurance in Manufacturing,'' cited to Emily Roberts and Rajesh Gupta, \emph{Journal of Industrial and Production Engineering}, $2021$.}
    {No bibliographic source returned a work matching the cited title or author pair.}
    {A deep web search surfaced no dedicated publication page, and Semantic Scholar, CrossRef, OpenAlex, DBLP, and the ACL Anthology held no record of the title.}
    {Retained as hallucinated: the reference is well formed and plausible, but it identifies no work that exists.}

The same pattern recurs across other references: a separate flag cited ``Multimodal Neural Machine Translation with Language Scene Graph'' to Lun~Huang et~al.\ at ACL~$2019$, and although our search surfaced thematically adjacent work on scene graphs and multimodal translation, none carried that title or those authors.

A second pattern keeps the title real but rebinds it to an author list that has nothing to do with the real authors; on a fast read the reference looks fine precisely because the title is genuine.

\exTPCard{Wrong authorship on a real title}
    {``Compressive Transformers for Long-Range Sequence Modelling,'' cited to D.~Dohan, X.~Song, A.~Gane, T.~Sarlos, et~al.}
    {The deep check found a paper with the same title, but its authors did not overlap with the cited list.}
    {Same title, but the real authors are Jack~W.~Rae, Anna Potapenko, Siddhant~M.~Jayakumar, Chloe Hillier, and Timothy~P.~Lillicrap.}
    {Retained as hallucinated: the title is real, but the cited authorship record is unrelated to the real one.}

\mypara{Why most flags are false positives.}
Equally important are the flags that look suspicious but resolve, on inspection, to real and recoverable works.
In many of these cases the flag is still reasonable, i.e., the automated check really did see a malformed author list or a metadata conflict. For example the following citations has a wrongly attributed authors (and sometimes titles), deciding if this is a hallucination or a human error is hard to determine but we choose to be on the more conservative side and consider it as a FP.

\exFPCard{Editor cited as a coauthor}
    {``Estimation of Non-Normalized Statistical Models by Score Matching,'' cited to Aapo Hyv\"arinen and Peter Dayan, \emph{Journal of Machine Learning Research}, $2005$.}
    {The cited list carried one more name than the matched record, so the automated check reported an author-count mismatch and even leaned toward a hallucination verdict.}
    {The JMLR paper is real, with the correct title, venue, and year; its sole author is Aapo Hyv\"arinen, and the additional cited name is the paper's editor rather than a coauthor.}
    {Set aside as false positive: the reference unmistakably identifies the real paper, and a wrongly attributed editor name does not break its scholarly identity.}

\exFPCard{Title fragment parsed as an author}
    {``A method for stochastic optimization,'' cited to D.~Kinga, Jimmy Ba Adam, et~al., at ICLR~$2015$.}
    {No source matched the truncated title and mangled author list, so the reference came back unverified with a hallucination-leaning verdict.}
    {The real paper is ``Adam: A Method for Stochastic Optimization'' by Diederik~P.~Kingma and Jimmy Ba (ICLR~$2015$); extraction dropped ``Adam:'' from the title and folded it into the author field as ``Jimmy Ba Adam.''}
    {Set aside as false positive: a parsing artifact split one of the field's most-cited references into a malformed title and author field, but the intended work is unmistakable.}

A subtler version arises when two genuinely distinct works share a title: an author-overlap check can lock onto the wrong one and report the citation as hallucinated.\footnote{Both appear in the SIGIR-AP~$2023$ accepted-tutorials list (Deng et~al; Balog and Zhai): \url{https://www.sigir-ap.org/sigir-ap-2023/accepted-papers/}.}

\exFPCard{Title collision between two real papers}
    {``Rethinking Conversational Agents in the Era of LLMs: Proactivity, Non-collaborativity, and Beyond,'' cited to Krisztian Balog and ChengXiang Zhai (SIGIR-AP~$2023$).}
    {Our tool matched the title to the better-indexed SIGIR-AP record by Yang Deng, Wenqiang Lei, Minlie Huang, and Tat-Seng Chua, whose authors do not overlap with the cited pair, the same signature as a wrong-authorship hallucination.}
    {SIGIR-AP~$2023$ in fact contains two distinct tutorials with this identical title: one by Deng et~al.\ and one by Balog and Zhai; the citation correctly identifies the latter.}
    {Set aside as false positive: both author lists are real for their respective papers, and the flag came from matching a same-titled twin rather than from a fabricated citation.}

\mypara{The hardest cases sit on the boundary.}
The clear cases are not what makes adjudication hard; deciding where the line falls is.
Author-list mismatches are the sharpest example.
How much divergence should count as a fabricated authorship record?
Disagreeing with the real list on half its names might sound like a reasonable cutoff, but ``half'' is fragile: on a two-author reference a single wrong name already crosses it, and one wrong name is exactly what an ordinary copy-and-paste slip can produce.
The kind of difference matters as much as its size.
A transposed or misspelled name is not the same as a genuinely different person, yet both register as a mismatch; adding the handling editor as if they were a coauthor, or promoting an affiliation string into the author list, inflates the discrepancy without inventing a work.
At this resolution, metadata alone almost never separates an erroneous  copy-and-paste error from a deliberate or machine-generated fabrication.
In this work we take a deliberately conservative approach to flagging hallucinated references, but even so, the tool's output is still a set of \emph{candidate} likely-hallucinated references that still require human verification and a human final decision. In other words, the tool filters out the verifiable references and leaves the unsure ones for a person to confirm.

\mypara{Verification fits naturally into the submission workflow.}
The practical implication is not a new gate that authors or reviewers must operate by hand, but a standardized check folded into existing submission infrastructure.
A venue can run our tool (or similar ones) automatically when a paper is submitted and/or at camera-ready, surface only the small set of references that fail verification, and attach the supporting evidence, candidate sources, matched metadata, and the deep-search trail, so that a human can confirm or dismiss each flag.
We propose this design out of respect for reviewers' workload: reviewing is voluntary and already strained by rising submission volumes, and resolving every reference in every bibliography by hand is neither a reasonable ask nor scalable.
By localizing suspicion to a handful of candidates and pre-attaching the evidence needed to judge them, automated verification converts an open-ended chore into a short, targeted confirmation step, leaving the final judgment with people while removing the part that does not scale.

\mypara{Author feedback suggests multiple failure modes.}
To further check our results, we contacted authors of accepted papers from a recent conference and shared the tool's output for their paper. We wrote only to papers with at least five likely hallucinated references. Across the responses, authors consistently attribute citation hallucinations to errors introduced during bibliography construction rather than to intentional fabrication. Multiple papers report that references were generated or reformatted using LLM-based tools, which produced incorrect or non-existent entries that were not subsequently verified, in some cases specifically during the camera-ready stage rather than the original submission. Other authors clarify that the cited works were real but contained corrupted metadata such as incorrect author lists, titles, or venues, while a smaller subset disputes certain flags as false positives (e.g., references to webpages or valid but misidentified papers). Where errors were acknowledged, authors uniformly report correcting the citations and, in several cases, proactively notifying program chairs or updating public versions. No response claims deliberate fabrication; instead, the explanations center on tooling-induced errors, metadata corruption, and insufficient manual validation in the citation pipeline.

Read together, these responses match our framing of the failure as a workflow breakdown, an unchecked, machine-completed reference that no stage of the pipeline was responsible for validating, rather than deliberate misconduct.

\mypara{Verification costs are almost negligible at venue scale.}
A recurring objection to pre-publication checking is cost, but at current prices the cost is negligible relative to the value of the archival record it protects.
In one venue-scale scan, auditing $3{,}703$ accepted papers and $221{,}281$ references cost roughly four cents per paper (\autoref{sec:results}).
At that price, verification can be run more than once, at submission to inform review, and again at camera-ready to catch references introduced or altered during revision, without materially affecting the budget of a conference.
We believe the dominant cost is not money but the human time needed to adjudicate the raised flags, which is precisely the quantity \tool{} is designed to minimize.

\mypara{Toward a shared citation-verification layer.}
Finally, our results point to a more durable fix than per-paper auditing.
Much of the friction we encounter, and a large share of our false positives, comes from the fact that bibliographic identity is fragmented across sources that disagree on titles, author lists, venues, and publication status.
A continuously updated, canonical citation service that unifies these records, built on or alongside existing aggregators such as Semantic Scholar, OpenAlex, CrossRef, DBLP, and the ACL Anthology, would help on both sides of the problem.
On the verification side, it would raise the precision and recall of any checker, including ours, by providing a single authoritative target to resolve against.
On the authoring side, it would let authors fetch correct, canonical bibliographic entries directly, reducing the chance that a partial or machine-completed reference is fabricated in the first place.
We believe improving citation integrity at the source, rather than only auditing it after publication, is a scalable way to keep the citation layer trustworthy as LLM-assisted writing becomes routine.

\section{Limitations}
\label{sec:limitations}

We now discuss several limitations for our study (and some of the reference hallucinations in general) and bound how its results should be interpreted and deployed.

\mypara{False positives are the primary limitation.}
Many candidate hallucinations are not hallucinations at all.
Of the flags we examined by hand, the majority were set aside as recoverable references (see the examples in \autoref{sec:discussion}).
We mitigate this with multi-source verification, deep-search re-verification, and by treating higher thresholds (at least two, and the five-or-more tail) as more reliable signals than any single flag.
Even so, an individual flag should never be treated as conclusive without review.
While false positives are a real problem, \tool{} still sharply reduces the number of references a reviewer must manually check.

\mypara{Results depend on the configured model, and so does cost.}
The reference-extraction and deep-search stages inherit the strengths and weaknesses of the configured LLM provider.
Stronger models tend to recover obscure sources and reject low-quality pages more reliably, reducing both false positives and false negatives, but they cost more per reference; weaker or cheaper models shift that trade-off in the other direction.
We used the best overall models for this study while keeping cost in perspective, namely Gemini 3.1 Flash Lite for reference extraction and Claude Haiku 4.5 for the deep-search stage.
We also experimented with upgrading the deep-search model to a Sonnet-class model, which substantially reduced false positives but increased per-reference cost significantly.
Finally, model behavior also changes over time (e.g., when providers roll out updated versions), so a scan run today may not reproduce exactly under a future model release.  

\mypara{Extraction errors are a leading cause of false positives.}
PDF reference extraction is the single noisiest step in the pipeline.
Merged or split fields, mangled author strings, truncated titles, and broken line structure can make a perfectly real reference look malformed, which is sometimes enough to trigger a likely-hallucinated verdict.
When authors or venues can supply the source bibliography directly, e.g., the \texttt{.bib} file, the pipeline can skip the noisiest stage entirely and verify cleaner inputs.
We recommend that venues collect source bibliographies at submission where possible, both to improve accuracy and to reduce the false positives created by extraction artifacts.

\mypara{Unpublished and not-yet-indexed works are near-certain false positives.}
Our verdicts are only as complete as the bibliographic record we resolve against.
Genuinely unpublished work, very recently accepted papers that the aggregators have not yet indexed, and obscure or non-archival venues can be impossible to confirm at scan time even though the cited work is real.
No tool can distinguish ``does not exist'' from ``not yet indexed'', so such references are systematically at risk of being flagged.
This is a structural reason to keep humans and author justification in the loop rather than acting on flags automatically.
We would like to highlight that this is not hypothetical: authors we contacted noted that some of their flagged references were for papers not yet published.

\mypara{The scope is the citation layer, not the paper's facts.}
We verify whether a cited work exists and whether its authorship record matches the citation.
We do not verify whether the citing sentence is faithfully supported by the cited work, nor whether the paper's scientific claims are correct.
A reference can resolve cleanly to a real, correctly attributed work and still be used to support a claim that the work does not actually make.
Citation-layer verification is thus necessary but not sufficient for paper integrity, and it complements rather than replaces claim-level fact-checking, which remains a harder problem.

\mypara{Desk rejection should leave room for author justification.}
We view recent policies that treat hallucinated references as grounds for desk rejection (\autoref{sec:related}) as a reasonable deterrent, and our results show the underlying signal is real and large enough to matter.
But even human judgment can be hard, for instance with not-yet-indexed works and genuinely ambiguous cases.
We therefore strongly recommend that flags inform an evidence-backed, human-reviewed process that gives authors an opportunity to justify or correct a flagged reference before any punitive action.

\section{Conclusion}
\label{sec:conclusion}

Hallucinated citations already appear in the proceedings of the field’s most selective venues, and almost nothing in the publication pipeline is built to stop them. That is the central finding of this work. At the level of individual references, the failure rate may look negligible (well under one percent in nearly every venue-year), but a conference is not one reference—it is hundreds of thousands. At that scale, the same rate yields an unverifiable citation in roughly a quarter of accepted NeurIPS~2025 papers, and our checks are nearly identical for accepted and rejected ICLR submissions whose reviewer scores differ by two full points.

We focus on references that cannot be verified, rather than citations with minor typos or metadata errors. Even then, it is difficult, even for humans, to distinguish a fabricated reference from a copy-and-paste slip, a wrong author from a misspelling, or a non-existent work from one that indexing services have not yet captured; no fixed similarity threshold can reliably make that call. What can be automated is the expensive part: resolving every reference against the bibliographic record, escalating the few that resist, and giving a reviewer evidence rather than a scavenger hunt. Our tool does this for cents per paper, and we open-source it so venues can verify the citation layer before publication.

\bibliographystyle{IEEEtran}
\bibliography{ref}

\end{document}